\documentclass{cimento}

\usepackage{graphicx}  
\title{Spectroscopical  Study of the Host Galaxy  of GRB031203}
\author{R.~Margutti\from{ins:x}\ETC,
C.~Chincarini\from{ins:x}\from{ins:evil},
D.~Fugazza\from{ins:evil}}
\instlist{\inst{ins:x} Universit\'a degli studi Milano Bicocca- Milano, Italy
  \inst{ins:evil} INAF - Brera Astronomical Observatory- Milano}
\PACSes{\PACSit{98.70.Rz}{Gamma-ray sources; gamma-ray bursts}%GRB, GRB Host Galaxies, Metallicity}
\PACSit{98.58.Ay}{Physical properties}
\PACSit{98.62.Ai}{Origin, formation, evolution, age, and star formation}}
\begin{document}
%98.58.Ay Physical properties (abundances, electron density, magnetic fields, scintillation, scattering, kinematics, dynamics, turbulence, etc.)  
%98.58.-w Interstellar medium (ISM) and nebulae in external galaxies  
%98.62.Ai Origin, formation, evolution, age, and star formation  
\maketitle

\begin{abstract}
We revisit the host galaxy of GRB031203 using a set of spectra obtained with VLT. Assuming a Galactic color excess $E(B-V) = 0.72 (mag)$
in the direction of the burst, we derive an internal extinction of $E(B-V)HG \approx 0.4 (mag)$.
After correcting for reddening, we find an electronic density of $156\pm21(cm^{-3})$ 
and a temperature of $12443\pm95 (K)$. With an ISM dominated by photo-ionization, we estimate a metallicity of 
$12+Log[O/H]=8.12\pm0.04$ and a star formation rate (SFR) 
of $12.3\pm2.1 (M_{o}yr^{-1}) $. This galaxy 
does not host a clearly detectable population of WR stars.
\end{abstract}

\section{Data Analysis and Discussion}

GRB031203 was detected by the IBIS instrument on board of the INTEGRAL satellite:
with a duration of about $30 (s)$ this burst is classified as a "long" GRB. 
GRB031203 was observed with ESO-VLT 
(Cerro Paranal, Chile) in five different epochs, with the last spectrum taken more than 9 
months after the explosion in order to be able to detect and follow the 
evolution of the Supernova and the possible evolution of parameters like 
the internal extinction for instance.
 The complete version of this work 
is presented in Margutti ~\cite{ref:1}. 
The photometry has been published in Malesani et al.~\cite{ref:2}

From the five strongest emission features of each spectra we derive a redshift 
$z=0.10536\pm0.00001$ which corresponds to a luminosity distance of  $473 (Mpc)$ when 
standard cosmology is used $(H_{o}=72$ ($Km$  $s^{-1} Mpc^{-1})$, $\Omega_{m}= 0.3$, $\Omega_{\Lambda }=0.7$ ):  this makes 
 GRB031203 one of the closest long burst ever observed. Because of the presence 
of the underlying supernova 2003lw,  HG031203 enter the class of galaxies of the
 local universe ($z \leq 0.1$) which are known to have hosted a LGRB and a  supernova 
explosion. In this work we compare HG031203 and other GRB-SNe Host galaxies (HG980425, HG030329, HG060218) 
to "normal" local galaxies of the universe.

The location of the burst close to the Galactic plane ($l = 255^{o}$,$b = -4.6^{o}$, Vaughan~\cite{ref:3}) 
makes very uncertain the amount of extinction correction due to Galactic absorption, because of the clumpy ISM. 
In order to have an estimate of this parameter we compared the reddening value from the Schlegel, 
Finkbeiner, Davies\cite{ref:4}  map scaled by a factor of $\approx30\%$ (Dutra et al.\cite{ref:5}) with the values of 
the closest galaxies available from literature and the reddening value of an object (ESO 314-2) 
symmetrical to  HG031203 with respect to the Galactic plane: we finally derive a Galactic color excess 
$E(B-V) = 0.72 (mag)$. We assessed the intrinsic extinction 
from Balmer decrements under the assumption of case B recombination (Osterbrock, Ferland \cite{ref:6})
and a Cardelli et al.\cite{ref:7} extinction law using $Rv = 3.1$ for both our Galaxy and the line of sight
through HG031203. We find $E(B-V)HG \approx  0.4 (mag)$. 

We derive an electronic density of $156\pm 21(cm^{-3} )$ and a temperature of 
$12443\pm 95 (K)$ using the standard recombination theory of [SII] and [OIII]. 
This results are similar to what was obtained by Prochaska et al.~\cite{ref:8}. In order to investigate the physical 
origin of the emission lines we compared the measured values of  [OIII]/H$\beta $ , [SII]/H$\alpha$, [NII]/H$\alpha$, 
[OI]/H$\alpha$ ratios of  HG031203 to those from samples of HII galaxies and AGNs from Osterbrock, 
Ferland,[6]. In particular, with  Log [[OIII]/H$\beta $] =0.81, Log[[SII]/H$\alpha$] =-1.21, Log[[NII]/ H$\alpha$]=-1.31,
Log[[OI]/H$\alpha$]=-2.02 ,  HG031203 is an object dominated by photo-ionization. We therefore exclude the 
presence of an AGN. 

Chemical abundances were estimated using the standard recombination theory
(Os~terbrock, Ferland ~\cite{ref:6}). Ionization correction factors by  Izotov, Thuan, 
Lipovetsky,~\cite{ref:9} were used in order to account for unobserved stages of ionization. 
We find that  HG031203 is characterized by low metallicity $(12+Log[O/H]=8.12\pm 0.04)$.
The same conclusion is supported by the comparison to the metallicity of  a large sample
of starbursting emission-line galaxies (ELGs) of the local universe (KISS galaxies, $z\leq 0.1$): 
according to the relation found by Melbourne, Salzer ~\cite{ref:10},  HG031203 would have 
had an oxygen abundance of $12+Log[O/H] \approx 8.9 $ which is above the value we derive (8.1). 
It's interesting to note that this doesn't seem to be a general feature of  HGs of LGRBs:
if we consider the other two galaxies of the local universe which are known to have hosted 
a correlated SN explosion and for which an estimate of metallicity is available from the
literature, then we find that  HG030329 has metallicity higher  than the mean relation
of the sample, while the properties of  HG980425 don't significantly differ from the 
reference sample. 

\begin{figure}[!h]
\begin{center}
\includegraphics[bb= 8 10 470 170 ,scale=0.74]{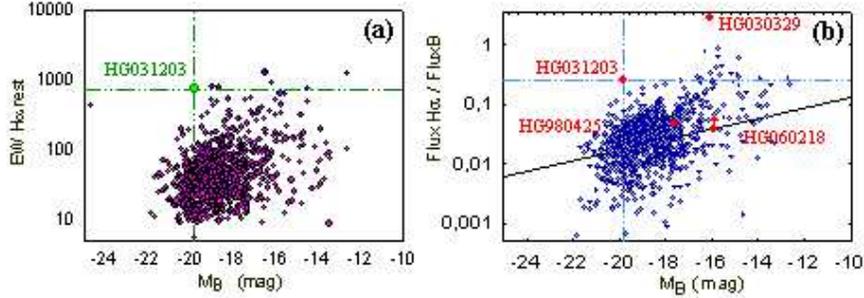}
\caption{ (a)EW H$\alpha$  at rest for  HG031203 and for a complete sample of 1029 local 
ELGs (KISS sample). (b) Specific star formation rate(SSFR) for galaxies of the local
 universe known to have hosted a LGRB and a correlated SN explosion compared to values
 of SSFR of the KISS sample. KISS data are not extinction corrected; SSFRs are then
 upper limits of the parameter. Thick line: regression of the sample. (Data about HGs taken 
from Sollerman,~\cite{ref:12}). GhostS database: www.pha.jhu.edu/~savaglio/ghost).}
\end{center}
\end{figure}

Using  H$\alpha $ emission, we estimate a $SFR$ of $12.3 \pm  2.1 (M_{o}yr^{-1})$,
in good agreement with the previous estimate of  Prochaska et al.,\cite{ref:8} 
($11 (M_{o}yr^{-1})  )$. HG031203 shows a prominent H$\alpha $ emission, with an equivalent width at
 rest of  750 (\AA) : it's clear from figure 1a that  HG031203 has an equivalent width H$\alpha $ at 
rest much greater than that of normal galaxies. In figure 1b, we compare  HG031203 to a 
complete sample of 1029 ELGs of the local universe (KISS galaxies, see Gronwall et al.~\cite{ref:11}
  for details about the sample). HGs of LGRBs-SNe are characterized by values of specific 
star formation rate (here defined as the ratio between  H$\alpha$   and B flux) higher than the mean 
of the reference sample as shown in figure 1b.

The progenitors of these LGRBs-SNe are likely WR stars.  
As shown in figure ~2 the identification of the 4686 HeII -the strongest
emission feature linked with WR  stars - is dubious (see however Hammer et al ~\cite{ref:13}).
This galaxy does not host a clearly detectable population of WR stars.\newline
\newline

\begin{figure}[!h]
\begin{center}
\includegraphics[bb= 0 10 450 170 ,scale=0.74]{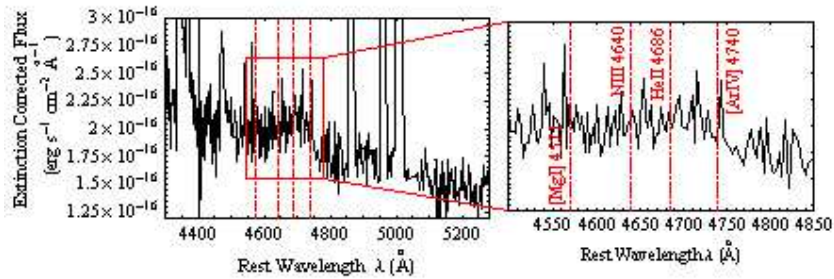}
\caption{ Extinction corrected flux of the HG031203. VLT observations taken on 18/09/04}
\end{center}
\end{figure}

\end{document}